\documentclass[a4paper,11pt]{article}
\usepackage{pos}
\newcommand{\bq}{\mbox{\boldmath $q$}}
\newcommand{\bp}{\mbox{\boldmath $p$}}

\title{Production of dileptons via photon-photon fusion in proton-proton
collisions with one forward proton measurement}
\ShortTitle{Production of dileptons in proton-proton
	collisions with forward proton}

\author*[a,b]{Antoni Szczurek}
\author[b]{Barbara Linek}
\author[b]{Marta {\L}uszczak}

\affiliation[a]{Institute of Nuclear Physics,\\
ul. Radzikowskiego 152, PL-31-342 Krak\'ow, Poland}

\affiliation[b]{College of Natural Sciences, Institute of Physics,\\
ul. Pigonia 1, PL-35-959 Rzesz\'ow, Poland}

\emailAdd{antoni.szczurek@ifj.edu.pl}

\abstract{We discuss the mechanism of dilepton production in
proton-proton scattering via fusion of virtual photons with 
identification of one proton on either side.
This is relevant for the ATLAS+AFP and CMS+PPS at the LHC.
Transverse momenta of the photons are taken into account via photon
unintegrated fluxes. The latter ones are expressed in terms of proton
electromagnetic form factors and structure functions.
We include different categories of such processes 
(double elastic, single dissociative). Some differential distributions
are shown explicitly and differences with respect to the results
without proton identification are discussed.
A soft gap survival factor is calculated using SuperChic-4 code.
}

\FullConference{%
  *** The European Physical Society Conference on High Energy Physics (EPS-HEP2021), ***\\
  *** 26-30 July 2021 ***\\
  *** Online conference, jointly organized by Universität Hamburg and the research center DESY ***
}

\begin{document}
\maketitle

\section{Introduction}

There are many mechanisms of dilepton production in proton-proton
collisions. One of them is photon-photon fusion mechanism.
So far such a mechanism was studied by measuring dileptons and
imposing a condition on rapidity gaps around dileptons.
Recently the CMS \cite{CMS} and ATLAS \cite{ATLAS} collaborations
presented results for the case when one proton is measured in forward
direction. We have developed a code which can calculate such processes
\cite{SFPSS2015,LSS2016,LSS2018}. In order to compare the theoretical 
results to the experimental data one has to impose kinematical condition
on so-called $\xi$-variable \cite{CMS,ATLAS}.

Here we show results based on our recent paper \cite{SLL2021}.
In our calculations we use the formalism developed by our group in
\cite{SFPSS2015,LSS2016,LSS2018}. How to include rapidity gap survival factor
related to emission of (mini)jets was disussed in
\cite{FLSS2019,LFSS2019}
for $W^+ W^-$ and $t \bar t$ production. In \cite{SLL2021} we repeated
similar analysis also for $\mu^+ \mu^-$ production.
The absorption for double-elastic contribution was studied e.g.
in \cite{LS2015,LS2018} within a momentum space formalism.
The impact parameter approach can be found in \cite{DS2015}.

\section{Sketch of the formalism}

\begin{figure}
\begin{center}
\includegraphics[width=4cm]{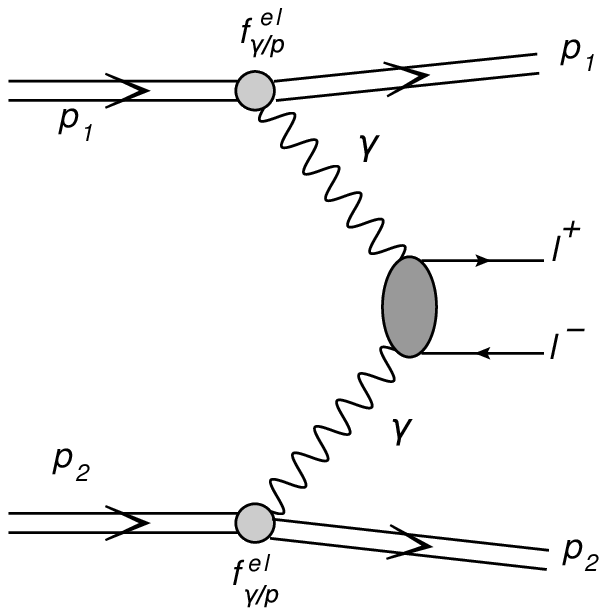}
\includegraphics[width=4cm]{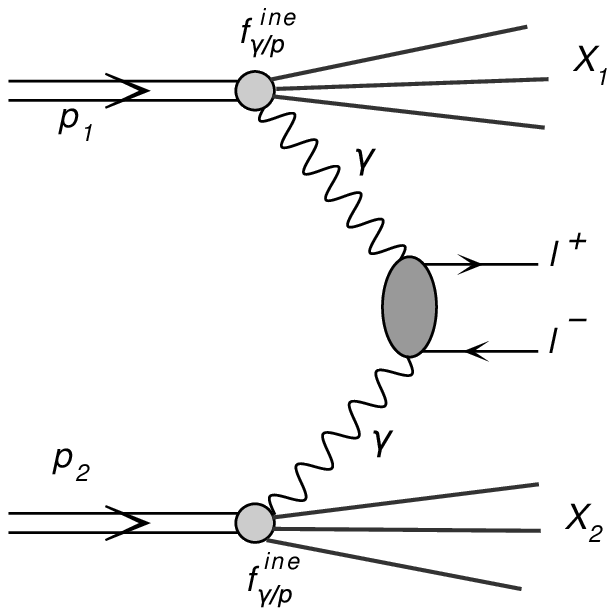}\\
\includegraphics[width=4cm]{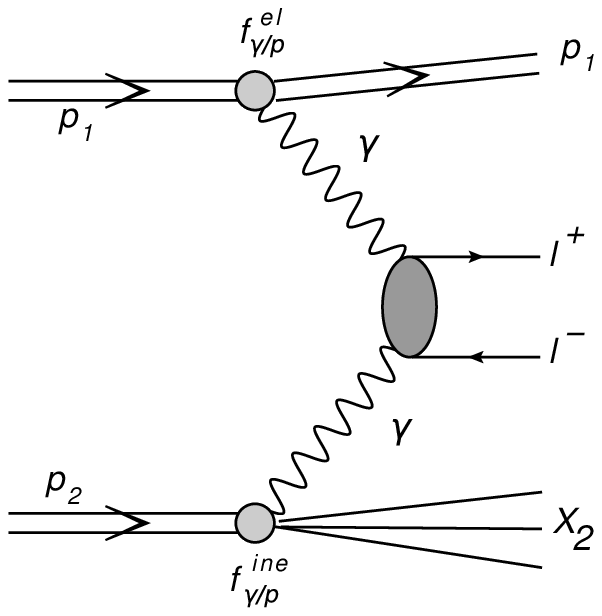}
\includegraphics[width=4cm]{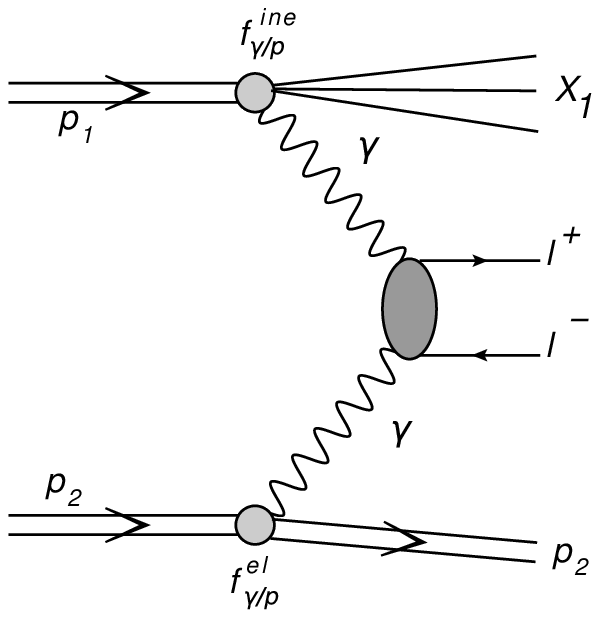}
\end{center}
\caption{Four different categories of $\gamma \gamma$ fusion mechanisms 
of dilepton production in proton-proton collisions.}
\label{fig:diagrams}
\end{figure}

There are four categories of the $\gamma \gamma$ processes
as shown in Fig.\ref{fig:diagrams}. We call them
elastic-elastic, inelastic-inelastic, elastic-inelastic and
inelastic-elastic. The double inelastic contribution is
not included when proton is measured (as there no explicit proton
appears in the formalism).

In the $k_T$-factorization approach  \cite{SFPSS2015,LSS2016}, 
the cross section for production of $l^+l^-$ can be written in the form
\begin{eqnarray}
{d \sigma^{(i,j)} \over dy_1 dy_2 d^2\bp_1 d^2\bp_2} &&=  
\int  {d^2 \bq_1 \over \pi \bq_1^2} {d^2 \bq_2 \over \pi \bq_2^2}  
{\cal{F}}^{(i)}_{\gamma^*/A}(x_1,\bq_1) \, {\cal{F}}^{(j)}_{\gamma^*/B}(x_2,\bq_2) 
{d \sigma^*(p_1,p_2;\bq_1,\bq_2) \over dy_1 dy_2 d^2\bp_1 d^2\bp_2} \, , \nonumber \\ 
\label{eq:kt-fact}
\end{eqnarray}
where the indices $i,j \in \{\rm{el}, \rm{in} \}$ denote elastic or 
inelastic final states.
Here the photon flux for inelastic case is integrated over the mass
of the remnant.

The ATLAS collaboration analysis imposes
a special condition on:
\begin{equation}
\xi_1 = \xi_{ll}^+  \; , \;  \xi_2 = \xi_{ll}^- \; .
\end{equation}
The longitudinal momentum fractions of the photons were calculated
in the ATLAS analysis as:
\begin{eqnarray}
\xi_{ll}^+ &=& \left( M_{ll}/\sqrt{s} \right) \exp(+Y_{ll}) \; , \nonumber \\
\xi_{ll}^- &=& \left( M_{ll}/\sqrt{s} \right) \exp(-Y_{ll}) \; .
\end{eqnarray}
Only lepton variables enter the formula.

\section{Results}

\begin{figure}
\begin{center}
\includegraphics[width=5cm]{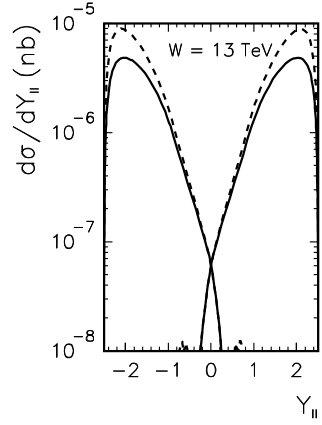}
\end{center}
\caption{Distribution in dilepton rapidity for
four different contributions considered.
Here the cuts on $\xi_{ll}^{+}$ or $\xi_{ll}^-$ are imposed.
The solid line is for double elastic contribution and the dashed line is
for single dissociation contribution.}
\label{fig:dsig_dYll_withcuts}
\end{figure}

As an example in Fig.\ref{fig:dsig_dYll_withcuts} we show a distribution
in $Y_{ll}$ (rapidity of the $l^+ l^-$ system). One can observe 
a deep dip at $Y_{ll} \approx$ 0 which is due to
the imposed cuts on the $\xi$-variable. When the cuts are removed 
the dip is not present \cite{SLL2021}.

\begin{figure}
\begin{center}
\includegraphics[width=5cm]{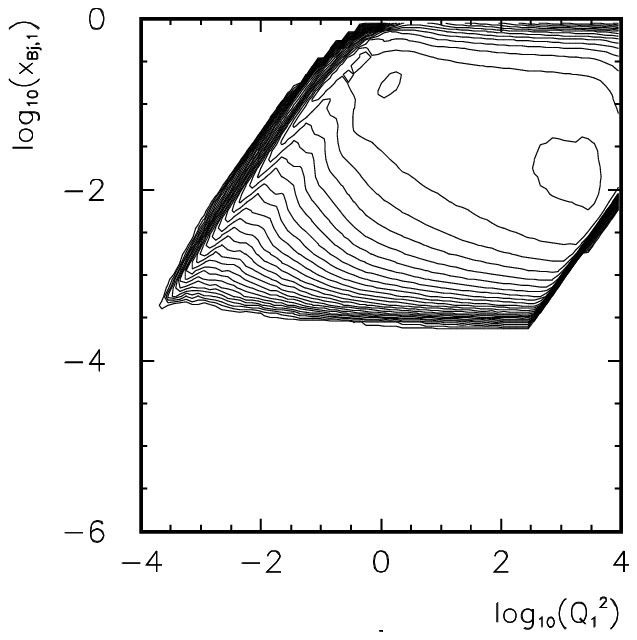}
\includegraphics[width=5cm]{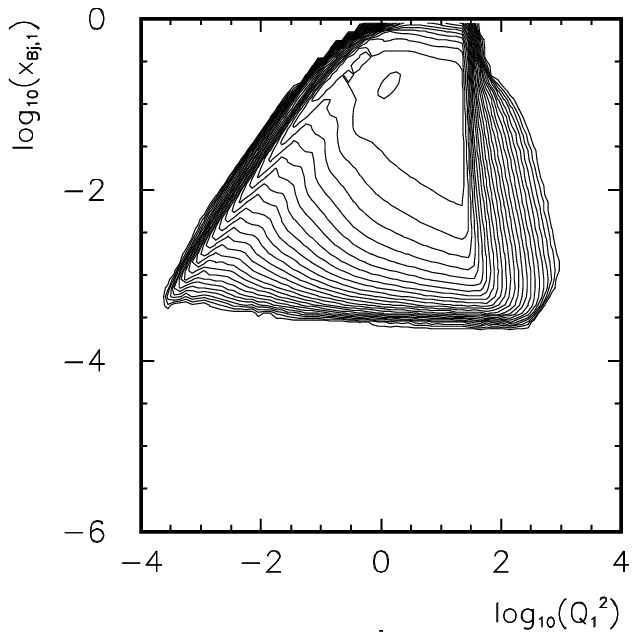}
\end{center}
\caption{
The range in the ($x_{Bj},Q^2$) space tested in
inelastic-elastic (left) with $\xi$ cuts. The right panel includes 
also an extra cut $p_{t,pair} <$ 5 GeV as imposed by the ATLAS
collaboration \cite{ATLAS}.}
\label{fig:xBj-Q2}
\end{figure}

What are typical $x_{Bj}$ and $Q^2$, i.e. arguments of the structure
functions for the considered processes with single proton dissociation,
is shown in Fig.\ref{fig:xBj-Q2}.
Both perturbative ($Q^2 >$ 2 GeV$^2$) and nonperturbative 
($Q^2 <$ 2 GeV$^2$) regions enter the corresponding calculations.
The nonperturbative region is even relatively larger
when the cut on $p_{t,pair} <$ 5 GeV is imposed as in the recent
ATLAS \cite{ATLAS} paper.

\begin{figure}
	\centering
\includegraphics[width=8cm]{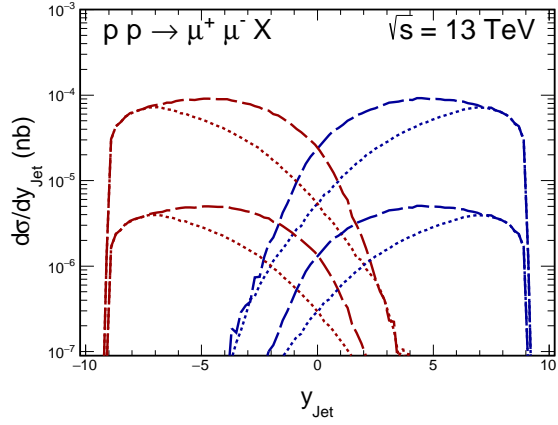}
\caption{Distribution in rapidity of (mini)jets for inclusive case
(upper curves) and for the case with cut on $\xi_{1/2}$ (lower curves).
The extra dotted lines represent results that include 
the cut $p_{t,pair} <$ 5 GeV as in the ATLAS analysis \cite{ATLAS}.}
\label{fig:dsig_dyjet}
\end{figure}

Many other distributions were discussed in \cite{SLL2021}.
The emission of photon coupled to quarks/antiquarks (single dissociative
process) may lead to 
a production of a extra minijet. Such s minijet leads to a distroying
rapidity gap and lowering the corresponding cross sections with gap
condition.
The rapidity distribution of such jets is shown in Fig.\ref{fig:dsig_dyjet}.

Now we wish to show some results obtained using the SuperChic-4
generator \cite{HTKR2020}.

\begin{figure}
\begin{center}
\includegraphics[width=6cm]{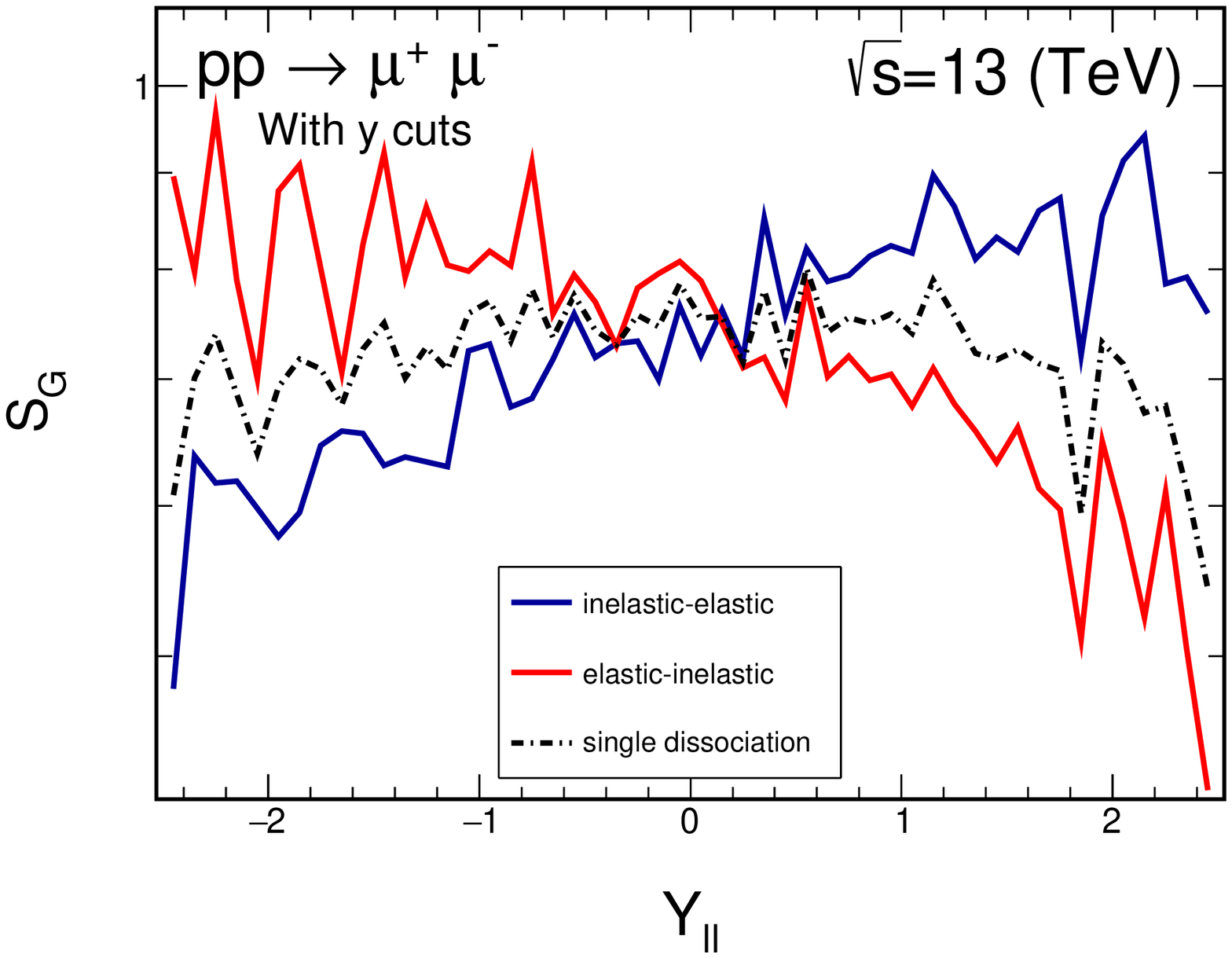}
\includegraphics[width=6cm]{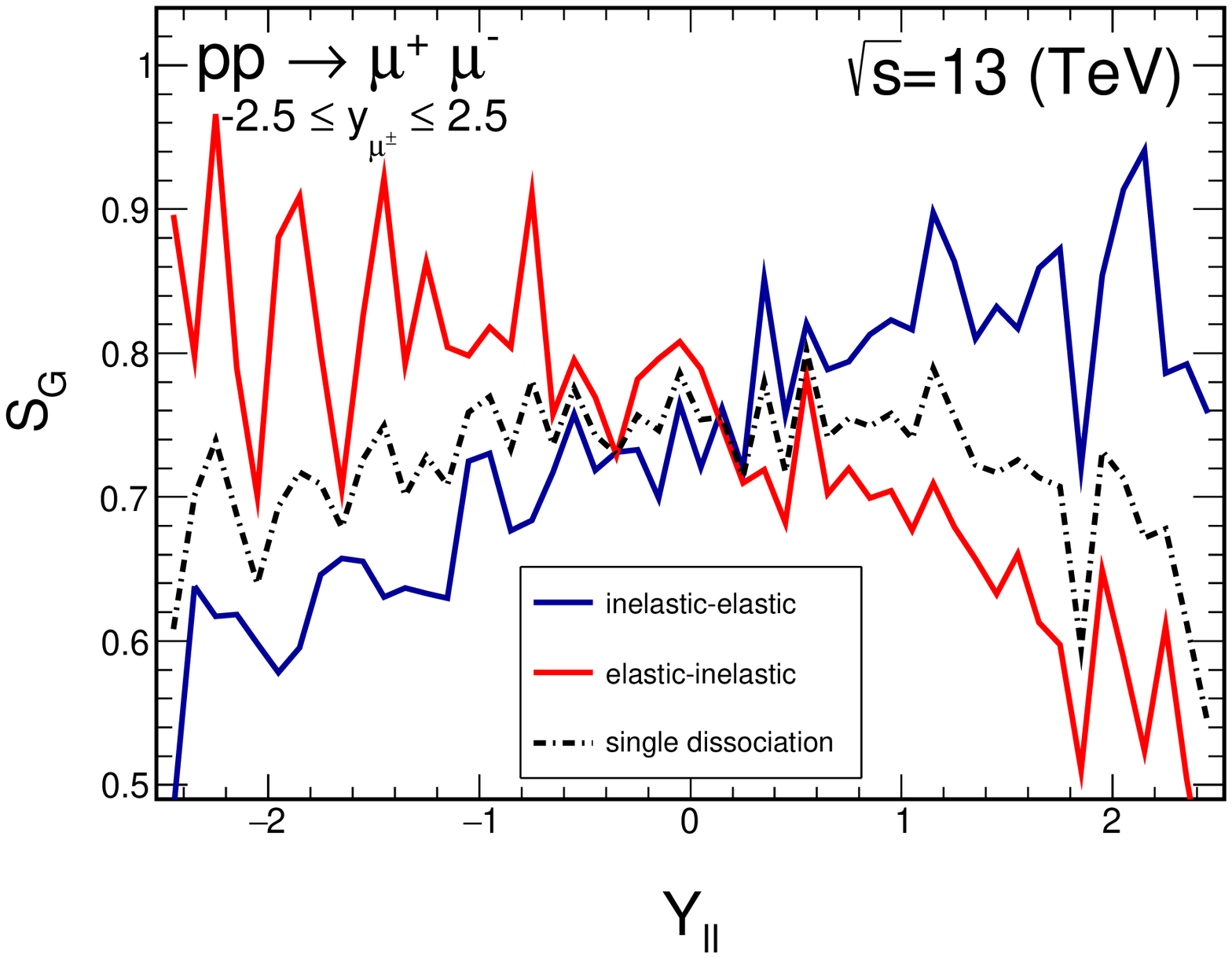}
\end{center}
\caption{The soft gap survival factor as a function of
rapidity of the $\mu^+ \mu^-$ pair for single proton dissociation.
We show the result without $\xi$ cuts (left panel) and
with $\xi$ cuts (right panel). The dash-dotted black line represents
effective gap survival factor for both single-dissociation components
added together.
}
\label{fig:soft_gap_survival_factor_2}
\end{figure}

In Fig.\ref{fig:soft_gap_survival_factor_2} we show corresponding
gap survival factor calculated as:
\begin{eqnarray}
S_G(Y_{ll}) &=& \frac{d \sigma / d Y_{ll}|_{with SR}}
                     {d \sigma / d Y_{ll}|_{without SR}}
\label{differential_gap}
\end{eqnarray}
as a function of $Y_{ll}$ variable.

Without the $\xi$ cut we observe quite different shapes of distributions
in $Y_{ll}$ without and with soft rapidity gap survival factor 
(see the left panel).
When the $\xi$-cut is imposed the distributions with and without
soft rapidity gap survival factor have very similar shapes.
Then, however, the elastic-inelastic and inelastic-elastic
contributions are well separated in $Y_{ll}$.

\begin{figure}
\begin{center}
\includegraphics[width=6cm]{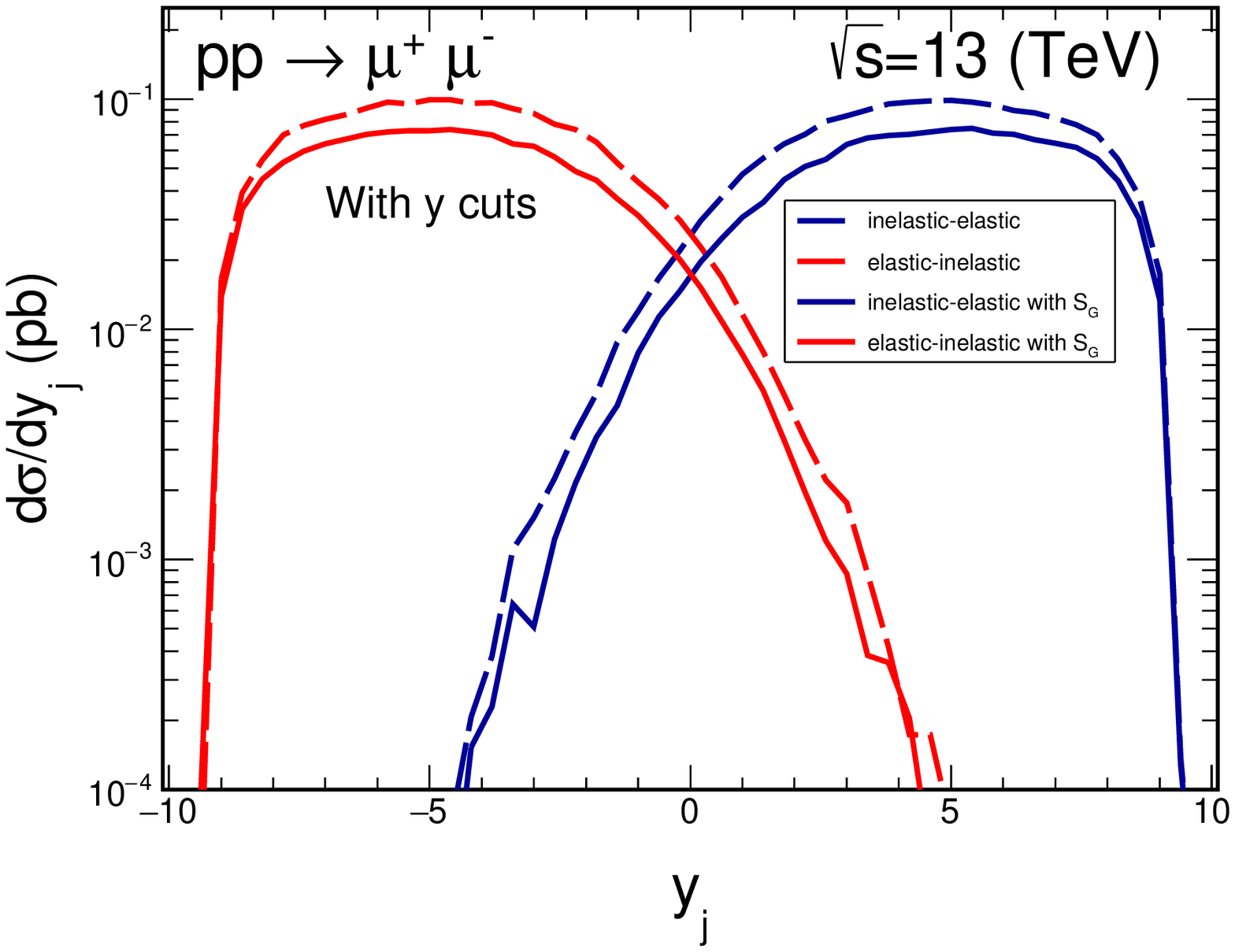}
\includegraphics[width=6cm]{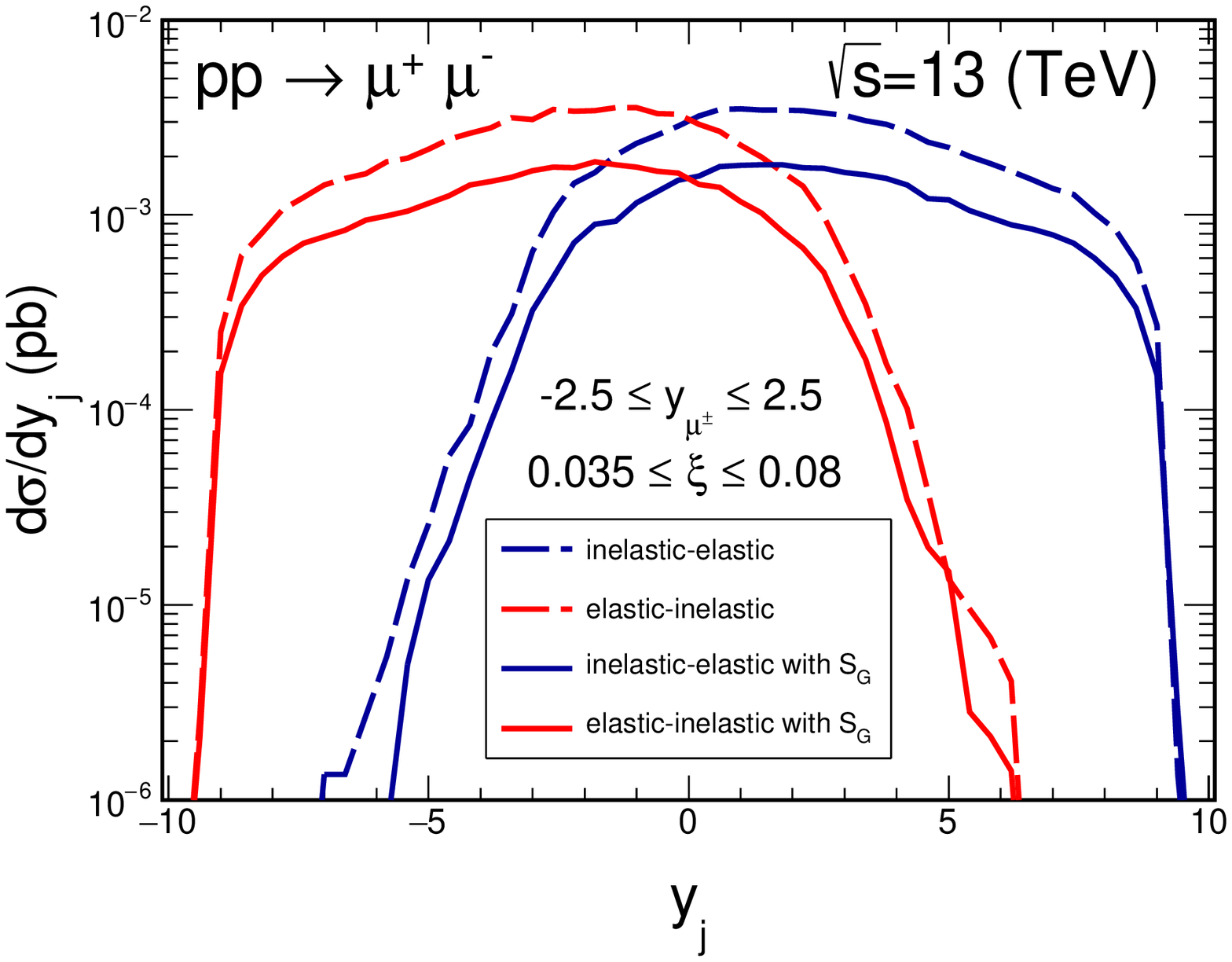}
\end{center}
\caption{Distribution in the (mini)jet rapidity for the inclusive case
with no $\xi$ cut (left panel) and when the cut on $\xi$ is imposed
(right panel) for elastic-inelastic and inelastic-elastic contributions
as obtained from the SuperChic-4 generator.
We show result without (dashed line) and with (solid line) soft
rescattering correction.
}
\label{fig:dsig_dyjet_SUPERCHIC}
\end{figure}

In Fig.\ref{fig:dsig_dyjet_SUPERCHIC} we show the (mini)jet distribution
in rapidity for elastic-inelastic and inelastic-elastic components.
We show the distribution without imposing the $\xi$ cut (left panel)
and when imposing the $\xi$ cut (right panel).
One can observe slightly different shape for both cases.
The corresponding gap survival factor (probability of no jet in the main
detector) is 0.8 and 0.5, respectively.

\section{Conclusions}

In this proceedings we have reported some selected results of our recent
studies of $\mu^+ \mu^-$ production in proton-proton scattering
associated with the emission of one forward proton.
This was accomplished by imposing a cut on proton energy loss fraction.
We have included both double elastic and single dissociative
contribution. The double-dissociative contribution was ignored
as here there is no explicit proton.
In our recent paper \cite{SLL2021} we considered both continuum 
production and production of $\Delta^+$ isobar and other resonances.

Several distributions were discussed in \cite{SLL2021}.
Here we have shown only some selected results.
Some difference compared to the case of no proton registration
conditions have been pointed out.
Particularly interesting is the distribution  
in $Y_{ll}$ which has a deep minimum at $Y_{ll} \sim$ 0.
The minimum at $Y_{ll}$ = 0 is caused by the application of a condition on 
$\xi_{ll}^{\pm}$ imposed on the leading proton.

For comparison we have shown results of calculation using popular
SuperChic-4 generator.
In general, the results are very similar to those obtained with
our codes.
We have shown also some results for kinematics-dependent
gap survival factor. We have found some interesting dependence
of gap survival factor on $Y_{ll}$.
Finally we have shown rapidity distribution of a (mini)jet associated
with partonic processes, also when including soft rescattering
corrections.

\section*{Acknowledgements}
This study was partially supported by the Polish National Science Center
grant UMO-2018/31/B/ST2/03537 and by the Center for Innovation and
Transfer of Natural Sciences and Engineering Knowledge in Rzesz{\'o}w.


\end{document}